\documentclass[a4paper,11pt]{article}
\pdfoutput=1 

\usepackage{jinstpub} 

\title{\boldmath Accurate GPS-based timestamp facility for Gran Sasso National Laboratory}

\author[a]{M. De Deo,}
\author[a]{G. Di Carlo,}
\author[a,c]{W. Fulgione,}
\author[a,b,1]{A. Molinario,\note{Corresponding author.}}
\author[a]{S. Parlati,}
\author[a,2]{R. Podviianiuk\note{Also at Institute for Nuclear Research, 03028, Kyiv, Ukraine.}}
\author[a]{and A. Razeto}

\affiliation[a]{INFN-Laboratori Nazionali del Gran Sasso,\\67100 L'Aquila, Italy}
\affiliation[b]{Gran Sasso Science Institute,\\67100 L'Aquila, Italy}
\affiliation[c]{INAF-Osservatorio Astrofisico di Torino,\\10125 Torino, Italy}

\emailAdd{andrea.molinario@lngs.infn.it}

\abstract{A new system to assign accurate timestamps to events recorded by experiments running underground has been designed, installed and tested at INFN Gran Sasso National Laboratory, Italy. This facility is based on a Master unit installed on surface and receiving time information from a GPS receiver, and Slave units placed underground which get data packet from the Master via optical fiber and assign the timestamps. The system is able to provide a time accuracy of 15 ns (1 $\sigma$) and precise reference frequencies to the experiments. It is now part of the infrastructure of the Laboratory for all the experimental activities which need accurate timestamps.}


\begin{document}
\maketitle
\flushbottom

\section{Introduction}
INFN Gran Sasso National Laboratory (LNGS), Italy, is the largest underground laboratory in the world. It is located between L'Aquila and Teramo, at about 120 kilometres from Rome, on one side of the 10-kilometer long highway tunnel which crosses the Gran Sasso massif, under a rock coverage of 1400 m (3600 m w.e.). It hosts various experiments in the field of direct dark matter detection, neutrino physics and nuclear astrophysics which benefit from the low cosmic ray background, with the muon flux reduced by a factor 10$^6$ with respect to the surface. 

During the life of the Laboratory, there has been the need to get a precise time reference in the underground experimental halls due to various physics case. A first example was the combined measurements of extensive air showers at the surface and high-energy muons deep underground~\cite{1990Bellotti,1992Aglietta}, which began in 1989 to end in 2000, involving MACRO~\cite{2002Macro} and LVD~\cite{1992LVD} experiments in the underground laboratory and EAS-TOP experiment~\cite{1989Aglietta}, the surface array, located above the Gran Sasso underground laboratory, at the altitude of 2005 m a.s.l. Since no physical connection existed between EAS-TOP and the underground laboratory, the correlation of data was established off-line, on the basis of the times of events recorded by the detectors. 
In second place, experiments aiming at the detection of neutrino bursts from supernovae in our Galaxy, like LVD and Borexino~\cite{2009Borexino}, need to be synchronized with other neutrino detectors in the world. This is in particular important to participate to the SuperNova Early Warning System (SNEWS)~\cite{2004Snews}, which has the goal to produce a prompt alert at the occurrence of a supernova looking at coincidence among neutrino detectors. A third case showing the need of accurate timing at LNGS is the measurement of neutrino from the CNGS neutrino beam~\cite{cngs1,cngs2} (2006-2012). The link between the neutrino production site at CERN and the detectors at LNGS was achieved with a common time reference. While the main goal of the project was the observation of $\nu_{\mu}\,\rightarrow\,\nu_{\tau}$ oscillations~\cite{2015Opera}, an accurate measurement of neutrino velocity was also performed by different experiments at LNGS: LVD~\cite{2012LVD}, ICARUS~\cite{2012Icarus}, Borexino~\cite{2012Borexino}, OPERA~\cite{2013Opera}. 

Any rare event search experiment currently running at the Laboratory or planned for the near future can benefit from an accurate timestamp facility at the occurrence of events which can be detected by detectors all around world, for example in the context of multi-messenger astronomy.  
It is also worth mentioning that such a system is able to provide, together with accurate timestamps of events, a standard reference frequency to the experiments.

The required accuracy for these different physics topics, with the exception of the neutrino velocity measurement, is of the order of few $\mu$s~\cite{1999Beacom,2009Pagliaroli}. Since it is not possible to directly receive a synchronization signal to the UTC time scale in the underground laboratory, any timestamp facility underground has to be linked to a station on surface. 

The first proposal of this kind of system dates back to 1987~\cite{1987Saavedra}. It was based on a Master unit (commercial device from ESAT company$^1$\note{http://www.esat.it}) located on surface in the external buildings of the Laboratory, close to the highway tunnel entrance, and on Slave clocks located underground. The Master was synchronized to the UTC time scale thanks to radio and TV reference signals, while Slaves received date, hour and synchronization from the Master via 8 km long optical fiber, connecting the surface and underground sites. These Slave clocks provided date and hour information to single experiments. This system has been running in LNGS for 30 years and it underwent some improvements during time. It is currently synchronized to the GPS timescale and it allows to date events in the experiments with an accuracy of 100 ns.


In occasion of the measurement of neutrino velocity with the CNGS neutrino beam (2012), in order to get the necessary higher time accuracy ($\sim$1 ns), a new dedicated system has been designed~\cite{2012Caccianiga} and installed in the external buildings of LNGS. A Septentrio PolaRx4 GNSS receiver, synchronized with the 10 MHz frequency of a GPS disciplinated Rubidium clock, was providing both the GPS time and a 10 Hz output signal. The system was equipped with high precision (50 ps) Time Interval Counters (TIC) Pendulum CNT-91, to which the triggers of the different LNGS experiments could be connected.
While this setup was designed and built with neutrino velocity measurement in mind, the new expertise acquired at the time and the availability of the new hardware motivated LNGS to move forward to a new version of the timestamp facility, to replace the old one. 

In section~\ref{sec:description} the general scheme of the new system is described, and then a detailed description of all its components is given. In section~\ref{sec:tests} the tests on the system and its performances are presented. Finally in section~\ref{sec:conclusions} the conclusions are given.

\section{Description of the system}\label{sec:description}

The new facility has been designed such that the LNGS will have control of the equipment, with the possibility to improve it and to produce the necessary number of units to serve the experiments which need it. For this reason custom-made components are used whenever possible. This approach grants a certain flexibility to match the different needs and requirements in the Laboratory. The design includes the continuous monitoring of all components in the system, to check they keep working properly and to promptly identify any misfunctioning. Since this new timing system is going to replace the previous one, the goal is to get an accuracy of the same order (100 ns) or better.
 
\begin{figure}[htp]
\centering
\includegraphics[scale=0.8]{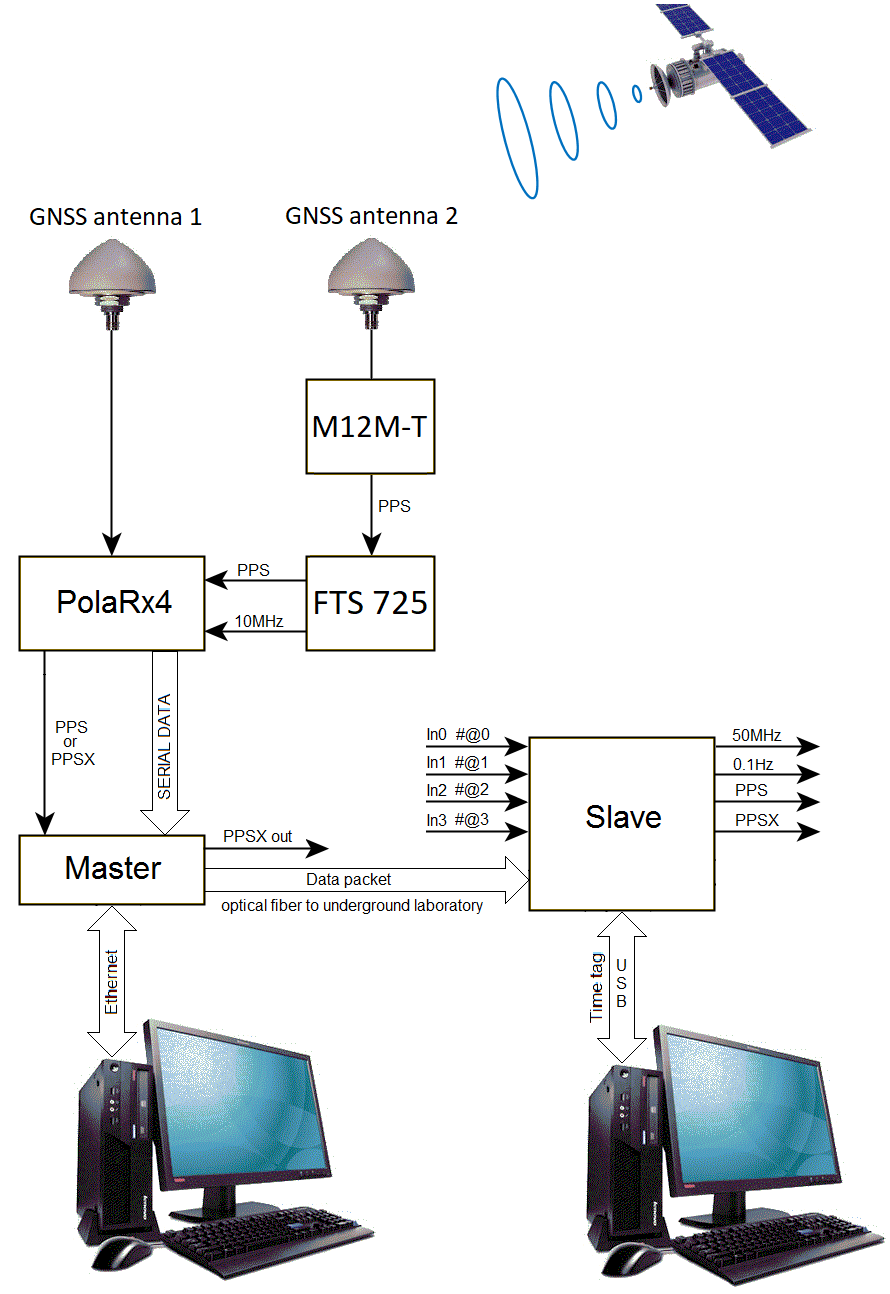}
\caption{The meaning of symbols is the following: M12M-T: single frequency/multichannel GPS receiver; FTS 725: GPS disciplined rubidium oscillator; PolaRx4: double frequency/multichannel GPS receiver. All these devices are located on surface, while the Slave is in the underground laboratory.}
\label{fig:schema1}
\end{figure}

The timing system includes three main units: GPS Receiver and Master on surface and Slave in the underground laboratory, see figure~\ref{fig:schema1}. The GPS Receiver consists of the commercial Septentrio PolaRx4, which receives the time information through GNSS antenna 1, coupled to a low jitter rubidium clock, see section~\ref{sec:receiver}. The GPS Receiver transmits the time information and a synchronization signal to the Master. The Master is a custom device, which is described in detail in section~\ref{sec:master}. It combines the synchronization signal and the time information from GPS Receiver in a single packet and transmits it to the Slaves underground via around 8 km optical fibers (section~\ref{sec:optical}). This is a simplex system where data transmits to only one direction from the Master to the Slaves. A Slave is a custom device, see section~\ref{sec:slave}, which uses the time information packet from the Master to assign a timestamp to any event produced by a detector underground.

\newpage

\subsection{GPS receiver and Rb}\label{sec:receiver}

The double frequency GPS Receiver Septentrio PolaRx4 is connected to a GNSS antenna (antenna 1). It receives the stream of data from GPS satellites and delivers a packet in binary format through RS232 serial connection at a rate of either 1/s (PPS) or 10/s (PPSX). The format of the output packet has been chosen to contain the informations regarding the quality of the signal received, the number of satellites available, the position and velocity of the receiver, the date and time as measured by the receiver and the difference of its local time with respect to the GPS time scale, the Clock Bias. Only part of this information is currently used to generate the final accurate timestamp, but it is possible to take advantage of the whole information to improve the quality of the system. A synchronization signal, which is generated with the same frequency of the packet, is sent out via coaxial cable. 

In order to provide the GPS receiver with a high stability time base, it receives both 10 MHz reference frequency and a pulse-per-second from a FTS 725 Rubidium Frequency Standard (Rb) from Stanford Research. The Rb oscillator is additionally locked with an auxiliary single frequency GPS receiver M12M-T by i-Lotus, connected to a second GNSS antenna (antenna 2). This configuration allows to continuously correct both the Rb oscillator frequency and the phase, providing a long term stability that is much better than the  one  achieved  with  the Rb oscillator only. The accuracy of the Rb oscillator in this configuration for 1 s windows has been measured at the time of neutrino velocity study at the Italian Institute of Metrology (INRIM), and resulted about 1.0$\times\,10^{-11}$ s/s~\cite{2012Caccianiga}.

\subsection{Master}\label{sec:master}

\begin{figure}[htp]
\centering
\includegraphics[scale=1.5]{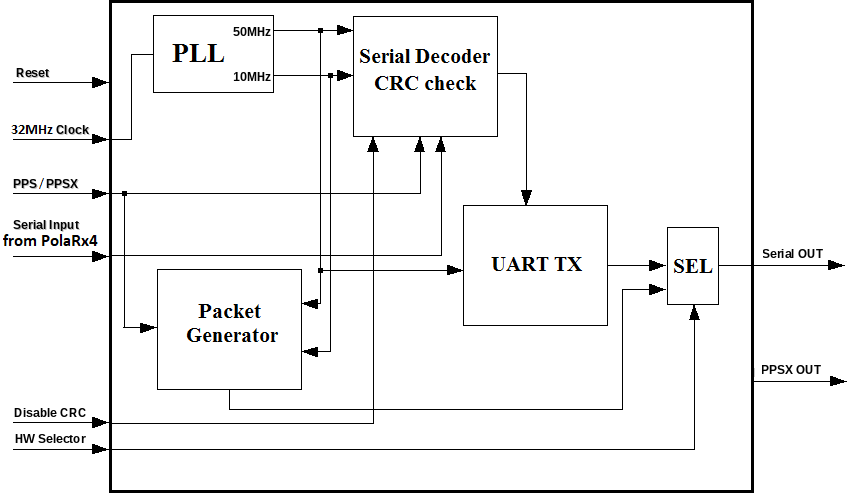}
\caption{Block diagram of the Master. The speed of data on Serial Input is 115.2 kbits/s, from Serial OUT is 10 kbits/s.}
\label{fig:schema2}
\end{figure}

The purpose of the Master is to collect and elaborate the serial data coming from the GPS Receiver to extract the necessary information, combine it with the synchronization signal and transmit this packet to the Slaves. The Master is based on a commercial FPGA development board with a customized firmware on it. The packet delivered to the Slaves contains a start bit, a Cyclic Redundancy Check with an error-detecting code (CRC), Coarse Time, i.e. the time since January 5, 2014 00:00:00 (UTC) in units of 0.1 s and the Clock Bias value. The block diagram of the Master is shown in figure~\ref{fig:schema2}. The 32 MHz clock signal is converted into 50 and 10 MHz by PLL (Phase Locked Loop). Clock signals with these frequencies go to the Serial Decoder and to the Packet Generator. The serial input packet from the PolaRx4  (section~\ref{sec:receiver}) is fed via Serial Input (115.2 kbits/s) to the Serial Decoder which extracts the time information from the packet and performs the CRC check. 

The decoded and verified data packet is sent to a UART TX (Universal Asynchronous Receiver-Transmitter) for the serial transmission to the Slaves at a rate of 10 kbits/s. The first bit of the transmitted packet is kept aligned to the PPS/PPSX transition from the GPS unit. In this way an accurate synchronization can be transmitted to the Slaves.


The Master is equipped with two low jitter J724 electrical/optical converters from Highland Technology which convert the serial output to optical signal which is sent to the Slaves via optical fibers. This setup allows to serve many Slaves at the same time, up to 16 in our configuration, with only one Master. Moreover, it grants a certain redundancy in case of failure of one of the two converters. It is also possible to read the Master output packet directly via USB and ethernet connections for the monitoring of the system.  

\subsection{Optical connections}\label{sec:optical}

The link between the Master on surface and the Slaves underground is performed via an optical fiber link. The electrical/optical converter acts as the signal transmitter (TX). The optical signal travels through approximately 8 km of a single-mode dedicated optical fiber to get underground. The signal can be split underground to serve different Slaves at the same time. The optical signal is converted back to electrical with a J730 optical/electrical converter from Highland Technology (signal receiver, RX), and fed to the Slave device. This converter is tuned so to properly match the signal strength after the attenuation due to optical path and to the splitter.

\begin{figure}[htp]
\centering 
\includegraphics[width=.35\textwidth]{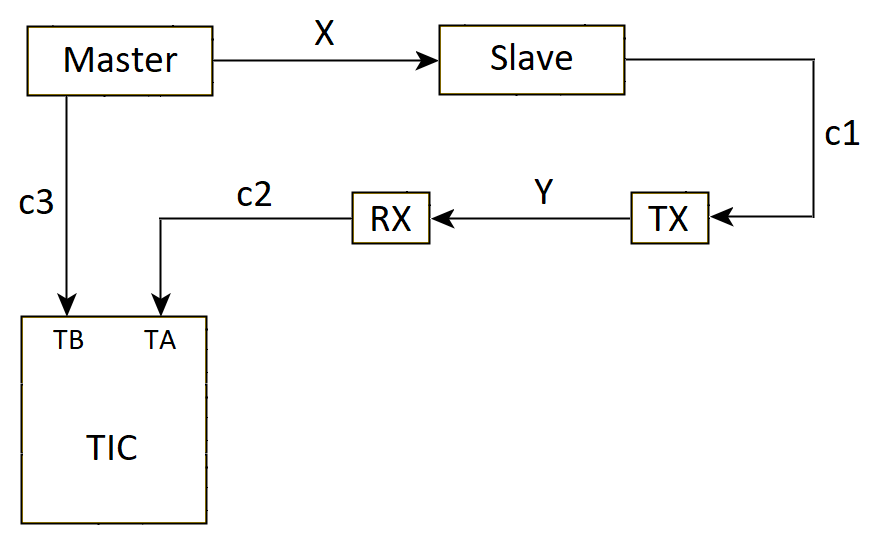}
\qquad
\includegraphics[width=.45\textwidth]{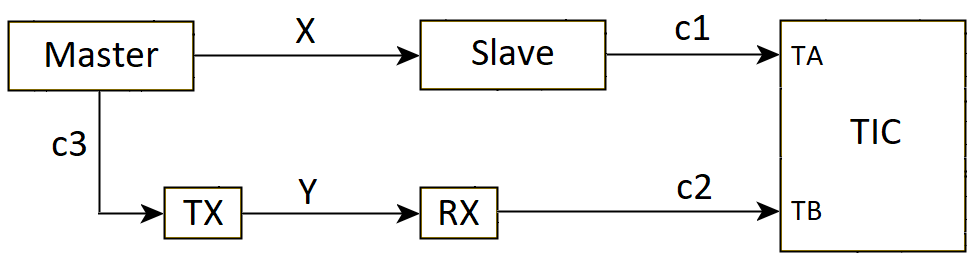}
\caption{\label{fig:fiber_setup} Setup for the measurement of the sum (left) and of the difference (right) of propagation times in two optical paths.}
\end{figure}

The time delay introduced by the optical fiber path is about 45 $\mu$s. It has to be measured individually for each Slave underground, with a precision of the order 1 ns. This is done building a second optical fiber path, and measuring both the sum and the difference of the signal propagation times along these two paths. The setups for these measurements are shown in figure~\ref{fig:fiber_setup}, where $X$ is the path to be measured, which starts at the Master input and ends at the Slave output, $Y$ is the second path, $c1$, $c2$ and $c3$ are cables, $TX$ is J724 electrical/optical converter, $RX$ is J730 optical/electrical converter, $TIC$ is an high precision (100 ps) time interval counter, which provides the time difference of two input signals $A$ and $B$ ($T_{A}\,-\,T_{B}$). A measurement of optical fiber time delay was performed with this method for the first Slave operating underground. It resulted to be $X\,=\,45977.0\,\pm\,1.2\,$ns. Periodic checks of the value of this delay will be performed.

\newpage
\subsection{Slave}\label{sec:slave}

\begin{figure}[htp]
\centering
\includegraphics[scale=1.4]{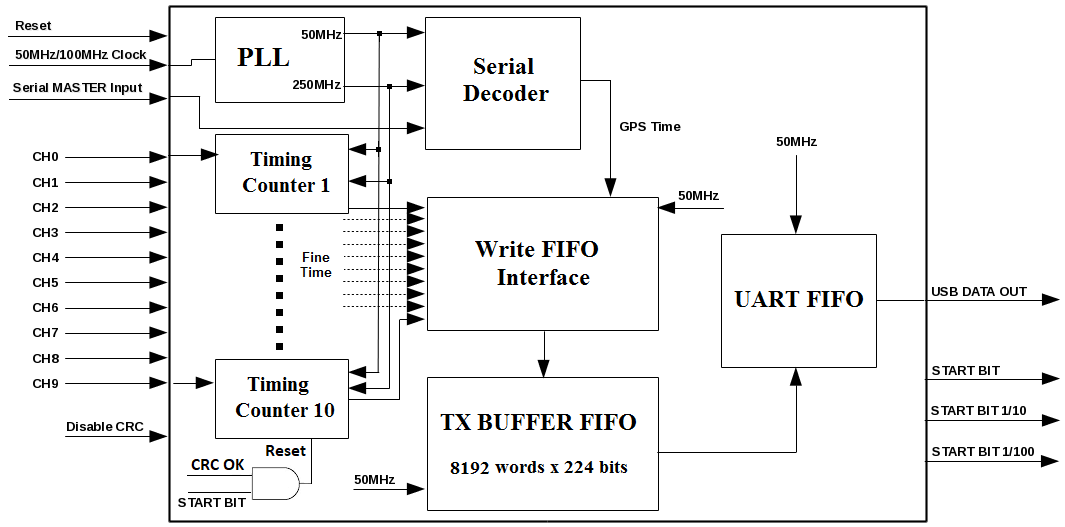}
\caption{Block diagram of the Slave.}
\label{fig:schema3}
\end{figure}

The purpose of the Slave is to collect the time information coming from the Master and assign accurate timestamps to signals coming to one or more of its input channels. Just like the Master, the Slave is based on a commercial FPGA development board with a customized firmware on it. The accurate timestamps is the sum of the Coarse Time (units of 0.1 s, see section~\ref{sec:master}) and of the Fine Time, which is the time between the arrival of the last packet from the Master and the signal on one input channel, measured with a precision of 4 ns. The number of input channels can be made up to 10 depending on needs of the user. The block diagram of the Slave is shown in figure~\ref{fig:schema3}. 

It contains, such as described for the Master in section~\ref{sec:master}, the PLL, Serial Decoder and UART blocks. An internal oven controlled crystal oscillator generates 50/100 MHz frequencies, which are converted into 50 and 250 MHz by the PLL. These frequencies are distributed to the Serial Decoder and Timing Counters. The data packet from Master comes to the Serial Master Input at PPS or PPSX frequency and it gets managed by the Serial Decoder, which writes the Coarse Time and Clock Bias values into the Write FIFO Interface. 

The Timing Counters produce the Fine Time by counting the number of 250 MHz clock cycles from the start bit of the packet received from the Master to the arrival of a signal (LVTTL logic) to one input channel (CH0-CH9). All Timing Counters get reset simultaneously when the Slave receives a new packet from the Master. The Write FIFO Interface produces the time record combining the number of the input channel which received the signal, the Clock Bias value, the Coarse Time and the Fine Time, and then writes this data into the memory buffer TX BUFFER FIFO. From this FIFO the data are transferred to the USB port from where they can be read by an external device. The Slave delivers precise reference frequencies reducing the data packet input frequency by different factors (x1, x1/10, x1/100).

\begin{figure}[hpt]
\centering
\includegraphics[scale=0.8]{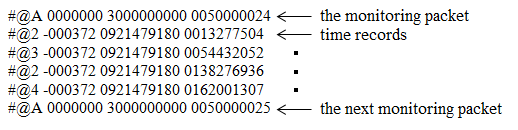}
\caption{Example of monitoring packets and time records as read from the Slave.}
\label{fig:records}
\end{figure}

Every time the Slave gets a start bit, a monitoring packet is also produced. It contains the number of 50 MHz clock cycles, $ClockCounts$, starting from the previous start bit. This information is used to monitor the drift $D$ of the internal oscillator of the Slave:\\
$D\,=\,(ClockCounts\,-\,ExpectedCountsValue)\,/\,(ExpectedCountsValue)$. \\
The monitoring packets are also sent to the TX BUFFER FIFO and read out via USB, at a speed of 1 Mbit/s. \\
The time information read from the Slave is organized in ASCII format and contains monitoring packets and time records, both of 34 bytes size. Given the reading rate, this imposes a limit of about 3.8 $\cdot 10^{3}$ monitoring packets and time records per second which can be written into the FIFO, above which it would eventually be filled up completely. An example of a series of monitoring packets and time records as read from the Slave is shown in figure~\ref{fig:records}.

The monitoring packet contains the following fields:
\\*
\textit{[Marker of the monitoring packet ] [Reserved] [Reserved] [Clock counts]}
\\*
where:
\\*
\textit{ [Marker of the monitoring packet]} This field contains \#@A, which identifies a monitoring packet.
\\*
\textit{ [Reserved]} Not used fields.   
\\*
\textit{ [Clock counts]} It is the number of 50 MHz clock cycles from the previous start bit.
\\*
\\*
The structure of the time record is:
\\*
\textit{[Number of the input channel] [Clock Bias] [Coarse Time] [Clock counts]}
\\*
where:
\\*
\textit{ [Number of the input channel]}  \#@0, \#@1, \#@2, ... \#@9 identifies which channel (0-9) received the input LVTTL signal.
\\*
\textit{ [Clock Bias]} This is the receiver Clock Bias with respect to GPS time system in ns, see section~\ref{sec:receiver}.
\\*
\textit{ [Coarse Time]} Time since January 5, 2014 00:00 (UTC), in units of 0.1 s. 
\\*
\textit{ [Clock counts]} Number of 250 MHz clock cycles from start bit to LVTTL signal on input channel. Fine Time is calculated multiplying Clock counts by 4 ns.

\subsection{Calculation of UTC time of events}\label{sec:timecalc}

It is possible to use the information contained in the Slave time record to calculate the number of nanoseconds that have elapsed since the Unix epoch (January 1, 1970 00:00 UTC), $T_{unix}$ (ns).\\

$T_{unix} (ns)\,=\,CoarseTime \cdot 10^{8}\,+\,FineTimeDC\,-\,CB\,+\,(T_{offset}\,-\,\Delta_{gpsutc}) \cdot 10^{9}\,+\,T_{fiber}$\\

$CoarseTime$ is described in section~\ref{sec:slave}. $FineTimeDC$ is the Fine Time (see section~\ref{sec:slave}) after correction to take into account the drift of the internal oscillator of the Slave. The drift $D$ can be calculated from the Clock Counts value in the monitoring packet as shown in section~\ref{sec:slave} , and the following relation holds: $FineTimeDC\,=\,FineTime\,/\,(1\,+\,D)$.
$CB$ is the Clock Bias of the receiver relative to GPS time system, see section~\ref{sec:receiver}. $T_{offset}$ is the number of seconds elapsed since the Unix epoch at January 5, 2014 00:00 UTC, which is the starting point of the time scale of the system. $\Delta_{gpsutc}$ is the difference between the GPS time scale and UTC time scale (+18 s as of 2018). $T_{fiber}$ is the delay introduced by the optical fiber path, which is measured as described in section~\ref{sec:optical}. Any user has also to properly include in this calculation all delays due to the propagation of the signal from the detector to the Slave, which are different for any particular experimental setup.\\


Two factors mainly determine the uncertainty of $T_{unix}$. One is the intrinsic accuracy of Septentrio receiver, affecting the $CoarseTime$ and $CB$ terms, which amounts to 10 ns (1$\sigma$)~\cite{septentriods}. This can be improved with off-line data processing taking into account ionosphere effects on the GPS signal propagation and proper geophysical models. In particular, the P3 and PPP (Precise Point Positioning) algorithms are commonly used for this purpose~\cite{2003Defraigne,2001Kouba}. The second factor is the uncertainty of the clock drift correction, which is at most 20 ns. The other terms in the expression for $T_{unix}$ are either constant ($T_{offset}$ and $\Delta_{gpsutc}$) or they have negligible uncertainty ($T_{fiber}$). The final total uncertainty on $T_{unix}$ amounts to about 15 ns (without P3 and PPP corrections). 



\section{Tests and performances}\label{sec:tests}
A series of different tests were carried out on the system to check its performances. In particular they were meant to measure the time accuracy, the stability and maximum input rate that the system can sustain. A troubleshooting campaign allowed to classify and understand the problems that arise when different parts of the system fail.

\subsection{Time accuracy}\label{sec:timeprec}
\begin{figure}[htp]
\centering
\includegraphics[scale=0.45] {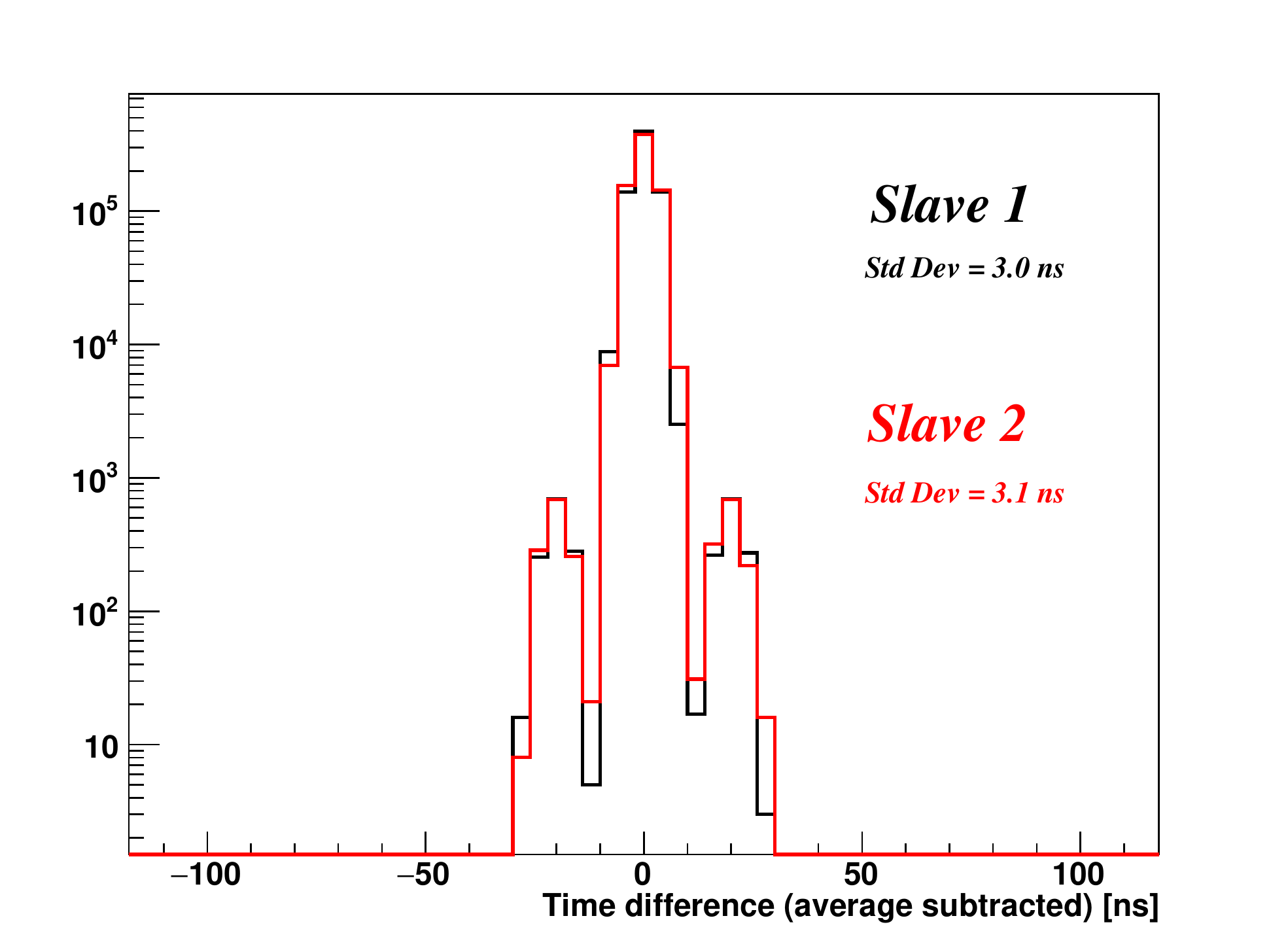}
\caption{Time difference between two consecutive pulses at fixed frequency in the first Slave (black line) and in the second one (red line), average subtracted.}
\label{fig:time_slave_both}
\end{figure}

A pulse of fixed frequency (1 Hz) has been generated by an external high stability Rb and sent to two different Slaves. The recorded timestamps have been corrected to take into account the drift of the internal oscillator of the Slaves. The resulting distribution of time differences between two consecutive pulses is shown in figure~\ref{fig:time_slave_both}, in black for first Slave and in red for the second one. The average value of the distribution has been subtracted for a better visualization. 
The time difference distributions for both Slaves have a standard deviation of about 3 ns. 
A population of events at $\pm$20 ns with respect to the expected value is apparent. These events happen at the same time in both Slaves, pointing to a common origin at the Master level. Each of these events is counted twice, once to calculate the difference with the previous event (ending in the -20 ns peak) and once in the difference with the following event (+20 ns peak). Further investigations on this effect are ongoing. However it is important to notice that it introduces a maximal shift of the measured times which stays within 30 ns.

\subsection{Stability}
\begin{figure}[htp]
\centering
\includegraphics[scale=0.5]{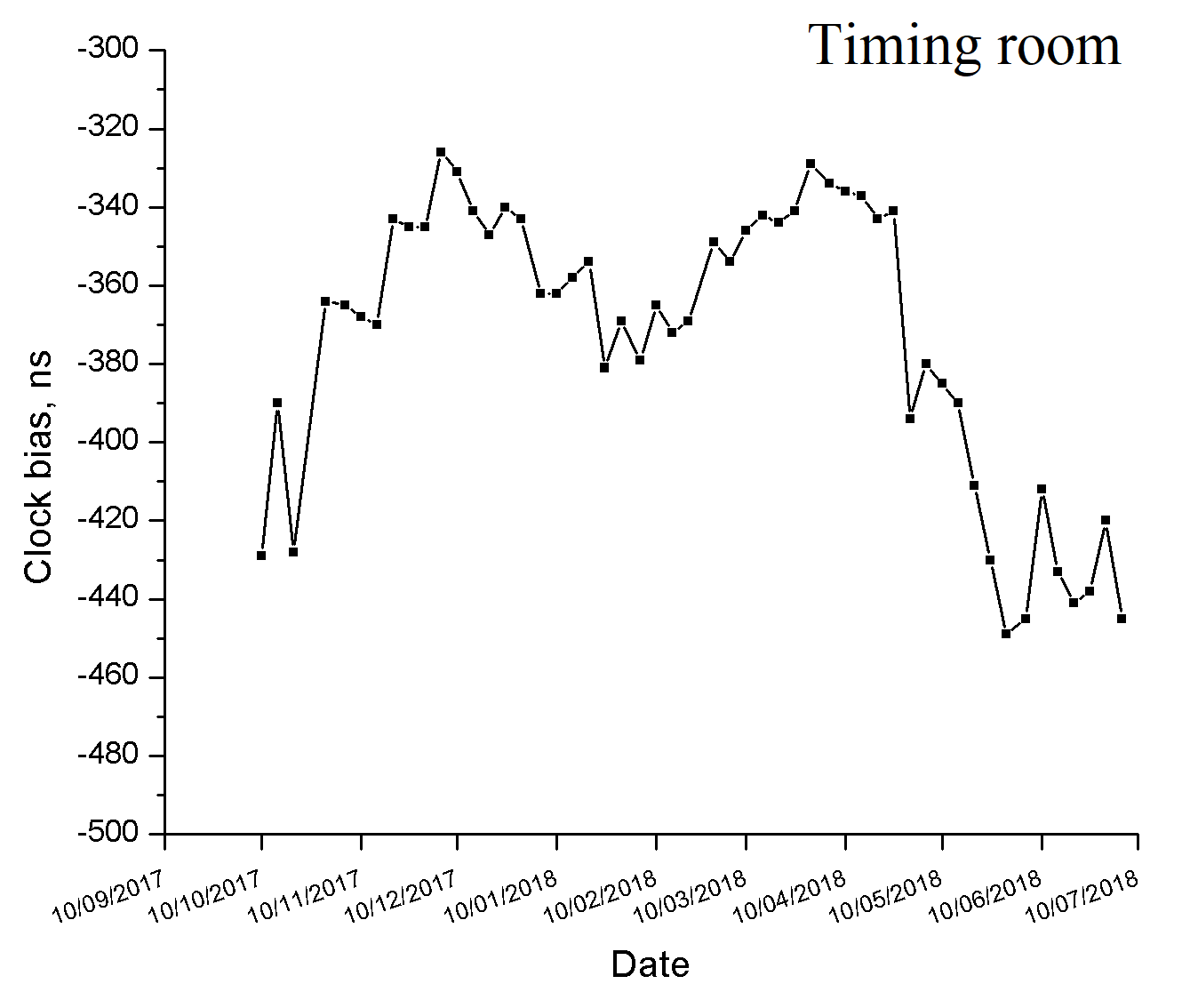}
\caption{Clock Bias evolution in time.}
\label{fig:cb_stability}
\end{figure}

\begin{figure}[htp]
\centering
\includegraphics[scale=0.5]{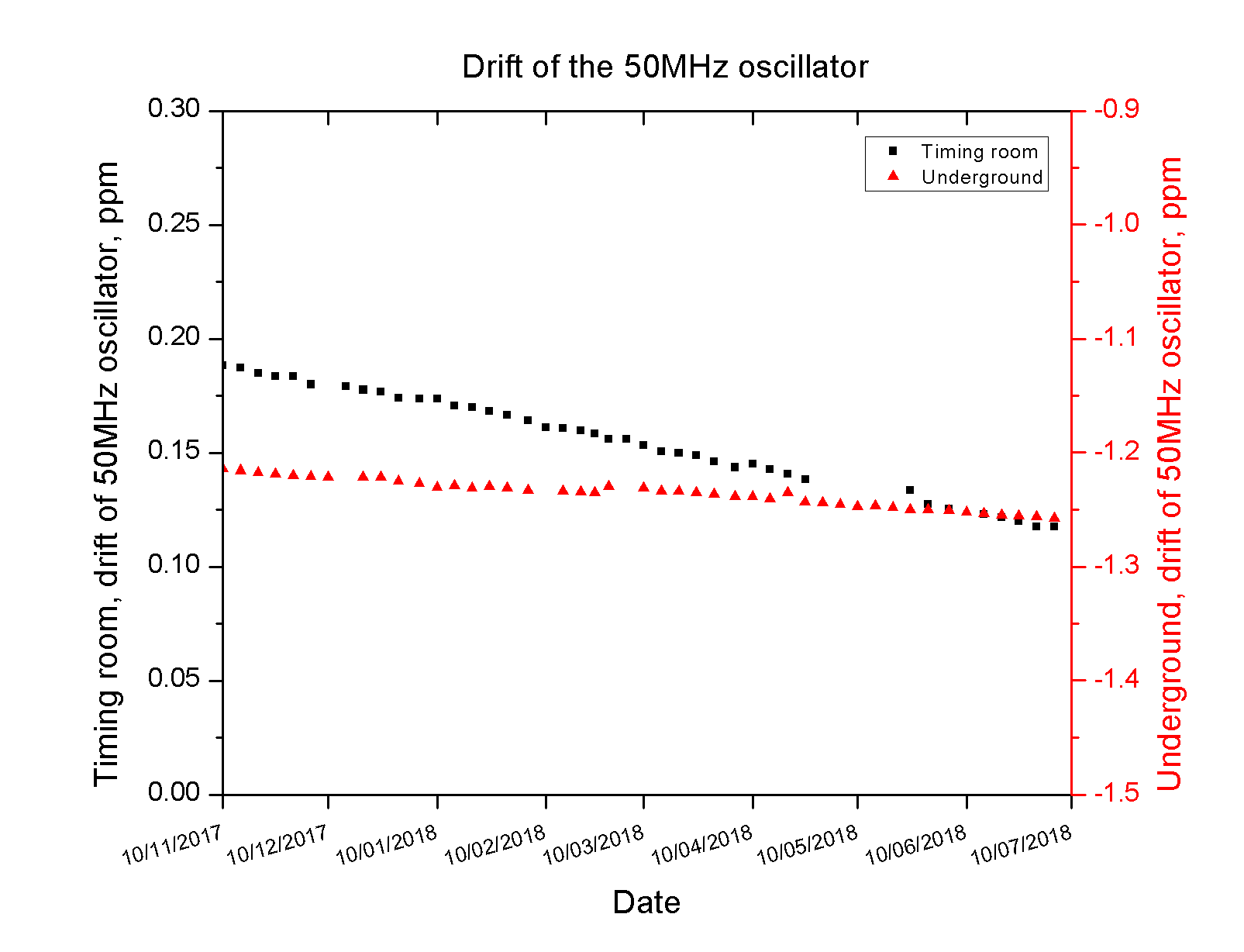}
\caption{Evolution in time of the drifts of internal oscillators for two different test Slaves.}
\label{fig:drift_stability}
\end{figure}

The stability of the system is continuously monitored by reading the Clock Bias value and measuring the drifts of the internal oscillators of the Slaves. Figure~\ref{fig:cb_stability} shows the evolution in time of the Clock Bias of the system, over a period of 9 months. The difference between the maximum and minimum values is 130 ns. Figure~\ref{fig:drift_stability} represents the evolution in time of the drifts of the internal oscillators of two different Slaves, over a period of 8 months. The initial value is $\sim$ 0.2 (-1.2) ppm for the first (second) Slave. As apparent from the figure, this value is slowly changing in time by -9$\cdot$10$^{-3}$ ppm/month (-5 $\cdot$10$^{-3}$ ppm/month) for the first (second) Slave. Both these effects can be taken into account and corrected with the information provided by the system itself.

\newpage
\subsection{Maximum input rate}

The maximum sustainable rate was measured by feeding to one channel an input signal of progressively higher rate (from 1 up to 20 kHz). The data were read via USB and written on a dedicated computer. It was found that 2.5 kHz is the maximum rate for which all signals are succesfully recorded by the system with proper measured time intervals between them. The test was repeated on a different channel, bringing the same result. The input rate was also distributed over different channels simultaneously, finding the same maximum value of 2.5 kHz considering the contribution of all input channels.

Beyond this value the system is still active and working, but it starts missing a fraction of the input signals since it saturates at the maximum rate. It is anyway important to recall that experiments running at LNGS are dedicated to the search of rare events, such that their typical event rate is much lower than the maximum input rate of 2.5 kHz. It is then expected that this is completely suited for the needs of present and future detectors at LNGS and in general for any experimental activity with an event rate not exceeding the limit of the system.\\

\subsection{Troubleshooting}
\begin{figure}[hpt]
\centering
\includegraphics[scale=1.3]{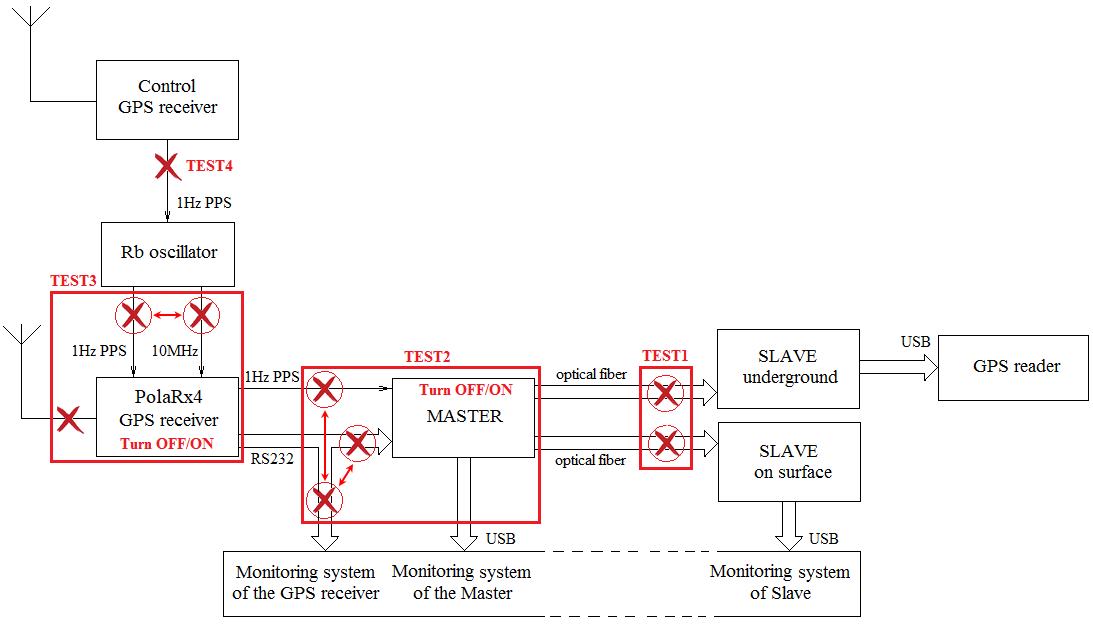}
\caption{Troubleshooting tests of the timing system.}
\label{fig:sch_troubleshooting}
\end{figure}

Troubleshooting is a systematic approach to problem solving that is often used to find and correct issues with complex machines, electronics, computers and software systems. Extensive tests have been performed on main units of the system and connections between them to check response of the system to situations when one or more failures, including communication problems, may happen. The system has points to check its status such as Monitoring system of the GPS receiver, Monitoring system of the Master, Monitoring systems of Slaves.
The scheme of the troubleshooting tests is shown in figure~\ref{fig:sch_troubleshooting}. 

There are 4 main steps (TEST 1-4) which include a total of 13 subtests. Each subtest introduces a known problem, in order to observe the reaction of the system to it. This allows in the future to identify which problem is affecting the timing system by checking its behaviour. 
\\*
\\*
\textit{ TEST 1 containing 2 subtests:}  Disconnection of the optical fibers connecting Master to Slaves
\\*
\textit{ TEST 2 containing 5 subtests:}  Operations on Master
\\*
\textit{ TEST 3 containing 5 subtests:}  Operations on Septentrio GPS receiver
\\*
\textit{ TEST 4:}  Disconnection of the control GPS receiver from the Rb oscillator 
\\*
\\*
The main symptom of almost half of subtests is a not changing value of the Coarse Time, the absence of the monitoring packet at the Slave and the saturation of Fine Time at its maximum value ($\sim$17 s). This can happen due to lost connection between Master and Slave, Master and GPS receiver or lost connection to the GNSS antenna. In this case it is necessary to check the entire sequence of connections to find where the failure happens. The other subtests show different symptoms which easily point to a specific problem. Thanks to the informations collected during the troubleshooting campaign it is possible to setup a monitoring system which can send out prompt alarm messages and warnings in case of failures.

\section{Conclusions}\label{sec:conclusions}
A new accurate GPS-based timestamp facility has been designed, assembled, installed and tested at LNGS. The goal of the system is to assign a precise timestamp to events in the underground laboratory to date them according to the UTC time scale and to provide standard reference frequencies to the experiments.

The time accuracy of the system, taking into account the uncertainties in the time calculation and the effects described in section \ref{sec:timeprec}, results to be about 15 ns (1 $\sigma$), well below the design goal of 100 ns. The stability of the system both underground and above ground has been measured and it is continuously monitored. The system can sustain a maximum rate of events to assign a timestamp of 2.5 kHz, which is completely suited for the typical experiments running at LNGS.  
A troubleshooting campaign has been carried out in order to predict and find solutions of most problems which can happen with system units and their connections on surface and underground. After commissioning phase, the timestamp facility became part of the infrastructure of LNGS, such that the experiments can benefit from it. An automatic monitoring and alarm system to allow prompt reaction in case of problems is under development.

\acknowledgments
We wish to thank the LNGS Director for supporting the project, N.Taborgna and the staff of Computing and Network Service for their valuable help in the installation of the new system. We appreciated the constant assistance and commitment of INTECS S.p.A., in particular M.Antonelli and M.R.Martinico, for the development of the dedicated firmware onboard of FPGA of Master and Slave. We are grateful to Borexino, Icarus and LVD collaborations for all the equipment we had at disposal, and to XENON collaboration for hosting the first Slave underground and for the support in the development of the firmware.


\end{document}